\documentclass{aa}  %% Astronomy & Astrophysics style class
\pdfoutput=1
\usepackage{graphicx,url,twoopt}
\usepackage[varg]{txfonts}           %% A&A font choice
\usepackage{hyperref}                %% for pdflatex
\makeatletter
\newcommand{\bibnote}[2]{\@namedef{#1note}{#2}}
\newcommand{\biblink}[2]{\@namedef{#1link}{#2}}
\makeatother

\begin{document}

\title{Detection of CI line emission towards the oxygen-rich AGB star omi Cet}
%  \\
%  \thanks{Research supported in part by the US Air Force
   % under grant no. AFOSR-88-0285 and
    %the National Science Foundation under grant
    %no. DMS-85-21154}\fnmsep
 % \thanks{This is a second footnote}\\
%  resulting in asymptotically faster convergence\\
%  for the same amount of work per iteration}
%\subtitle{H$^{12}$CN and H$^{13}$CN excitation analysis in thecircumstellar outflow of R Scl\\}
% \author{M. Saberi\inst{1}
\author{M. Saberi \inst{\ref{inst1}}
      \and W. H. T. Vlemmings\inst{\ref{inst1}}
      \and  E. De Beck\inst{\ref{inst1}}
      \and R. Montez\inst{\ref{inst2}}
      \and S. Ramstedt\inst{\ref{inst3}}
      }
\institute{Dept. of Space, Earth and Environment, Chalmers University of Technology, Onsala Space Observatory, 43992 Onsala, Sweden \email{maryam.saberi@chalmers.se}\label{inst1}
\and Smithsonian Astrophysical Observatory, Cambridge, MA 02138, USA\label{inst2}
\and Department of Physics and Astronomy, Uppsala University, Box 516, SE-75120, Uppsala, Sweden\label{inst3}}
%\and \label{1} {M. Saberi}\inst{\ref{1}}
% {Department of Physics and Astronomy, Institute of Astronomy, KU Leuven, Celestijnenlaan 200D,  3001 Leuven, Belgium}\label{2}
%}
\date{}

\abstract 
%{Context.}{Methods.}{Results.} {Aims. }
{We present  the detection of neutral atomic carbon CI($^3P_1$ - $^3P_0$) line emission towards omi Cet. This is the first time that CI is detected in the envelope around an oxygen-rich M-type asymptotic giant branch (AGB) star. 
We also confirm the previously tentative CI detection around V Hya, a carbon-rich AGB star. 
As one of the main photodissociation products of parent species in the circumstellar envelope (CSE) around evolved stars, CI can be used to trace sources of ultraviolet (UV) radiation in CSEs. 
The observed flux density towards omi Cet can be reproduced by a shell with a peak atomic fractional abundance of 2.4 $\times$ 10$^{-5}$ predicted based on a simple chemical model where CO is dissociated by the interstellar radiation field. However, the CI emission is shifted by $\sim$ 4 km s$^{-1}$ from the stellar velocity. 
Based on this velocity shift, we suggest that the detected CI emission towards omi Cet potentially arises from a compact region near its hot binary companion. The velocity shift could, therefore, be the result of the orbital velocity of the binary companion around omi Cet.
In this case, the CI column density is estimated to be 1.1 $\times 10^{19}$ cm$^{-2}$. 
This would imply that strong UV radiation from the companion and/or accretion of matter between two stars is most likely the origin of the CI enhancement. 
However, this hypothesis can be confirmed by high-angular resolution observations.}

\keywords{Stars: abundances -- Stars: AGB -- Stars: individual: omi Cet, V Hya --  Stars: binaries -- Stars: carbon -- Stars: circumstellar matter -- Stars: chromospheres}
\maketitle
%A&A Editorial Office: Astronomy & Astrophysics - Author?s guide 21

%%%%%%%%%%%%%%%%%%%%%%%%%%%%%%%%%%%%%%%%%%%%%%%%%%%%%%%%%%%%%
\section{Introduction}\label{Introduction}
%%%%%%%%%%%%%%%%%%%%%%%%%%%%%%%%%%%%%%%%%%%%%%%%%%%%%%%%%%%%%

Low- to intermediate-mass stars lose a substantial amount of their mass through strong winds during the asymptotic giant branch (AGB) phases of stellar evolution. As a consequence, a large circumstellar envelope (CSE) containing gas and dust will form around the star \citep{Habing96}.
These large CSEs around evolved stars provide a unique laboratory to study the late stellar evolutionary phases.

Destruction of carbon-bearing molecules by either ultraviolet (UV) photo-dissociation or shock-dissociation leads to the possible presence of CI in these CSEs \citep[e.g.][]{Glassgold86}. An enhancement of the CI line emission from these environments can, therefore, probe UV- and shock-induced chemistry.

Previous studies show the CI/CO ratio increases significantly as an AGB star evolves to the post-AGB and planetary nebula phases, suggesting an evolutionary sequence for the CI/CO ratio \citep[e.g.][]{Bachiller94, Young97, Knapp00}.
These studies suggested that CI initially forms by UV photodissociation of C-bearing molecules due to the interstellar radiation field (ISRF) when the star is at the onset of the AGB. At the end of the stellar evolution, the ratio increases significantly due to the additional photodissociation in the inner envelope by the hot central star in the post-AGB and planetary nebula phases \citep[e.g.][]{Knapp00}.

To date, CI has been observed around several planetary nebulae which still contain part of their envelopes. CI detections are also reported for four post-AGB stars \citep[][and references therein]{Knapp00} and C-type AGB stars IRC+10216 \citep{Keene93, vanderveen98}, R Scl \citep{Olofsson15} and (tentatively) V Hya \citep{Knapp00}. 
For the AGB stars IRC+10216 and R Scl, CI appears in shells, implying that the ISRF is the main source of CI formation.

Although penetration of UV radiation from the ISRF has been considered as the main source of UV radiation in CSEs around AGB stars, there is ample evidence for the presence of internal UV radiation in the CSEs. The internal UV radiation can be generated by a hot binary companion, accretion of matter between two stars, and stellar chromospheric activity \citep[e.g.][]{Sahai08, Linsky17}. Recent Galaxy Evolution Explorer (GALEX) observations revealed 180 AGB stars (57 $\%$ of the observed sample) with detectable Far- and/or Near-UV emissions \citep{Montez17}, proving the presence of the internal UV radiation.

To address the effects of both internal and external UV radiation sources, observations of the main photodissociation/photoionization products from the most abundant species such as CO are required.  
Chemical models of CSEs that consider both an internal and an external UV radiation field predict the enhancement of CI and/or CII in the inner CSE (Saberi et al. in prep). The radial distribution and the peak abundance of CI and CII mostly depend on the strength of the UV field and the H$_2$ density in the CSE. Observations of CI and CII in AGB stars with strong UV detection will help to further constrain the chemical modelling.

%%%%%%%%%%%%%%%%%%%%%%%%%%%%%%%%%%%%%%%%%%%%%%%%%%%%%%%%%%%%%
\section{Sources}\label{Sources}
%%%%%%%%%%%%%%%%%%%%%%%%%%%%%%%%%%%%%%%%%%%%%%%%%%%%%%%%%%%%%

%%%%%%%%%%%%%%%%%%%%%%%%%%%%
\subsection{omi Cet}
%%%%%%%%%%%%%%%%%%%%%%%%%%%%

omi Cet belongs to the closest symbiotic binary system, the Mira AB system, which is reported to be at a distance of 92 pc \citep{vanLeeuwen07}.
The primary star omi Cet (Mira A) is an M-type AGB star with mass loss of 2.5 $\times$ 10$^{-7}$ M$_\sun$ yr$^{-1}$ \citep[][hereafter RS01]{Ryde01}, and the companion, VZ Cet (Mira B) is believed to be a white dwarf \citep{Sokoloski10}. 
Previous imaging of the Mira AB system has shown that the circumstellar material of omi Cet is flowing towards the companion \citep{Karovska97,Karovska05}.
Moreover, \cite{Ramstedt14} observed a bubble-like structure in the south-east part of the Mira AB system in CO(3-2) line emission. They suggested that it is formed due to blowing of the circumstellar material of omi Cet by the wind of the companion (Mira B). 
They also show that the CO emission is very extended ($\sim$20") and arises in a range of local-standard-of-rest (lsr) velocities 37 to 54 km s$^{-1}$ which indicates a relatively low wind velocity of 5 km s$^{-1}$. However, this velocity is larger than the 2.5 km s$^{-1}$ wind velocity reported by RS01.

UV emission from the Mira AB system has been studied by numerous space-based UV observatories \citep{Cassatella79,Stickland82,Karovska97,Wood01,Wood04,Martin07}.
Broad band fluxes from more recent UV observations with the GALEX are reported in \cite{Montez17}.

%%%%%%%%%%%%%%%%%%%%%%%%%%%%%%%%%%%%%%%%%%%%%%%%%%%%%%%%%%%%%
\subsection{V Hya}
%%%%%%%%%%%%%%%%%%%%%%%%%%%%%%%%%%%%%%%%%%%%%%%%%%%%%%%%%%%%%

V Hya is a carbon-type AGB star which is believed to be in transition to the planetary nebulae phase \citep[e.g.][]{Knapp00,Sahai16}. It is located at a distance of 380 pc \citep{Perryman97} and has a mass loss rate of 1.5 $\times$ 10$^{-6}$ M$_\sun$ yr$^{-1}$ \citep{Knapp00}.
\cite{Sahai16} reported a high-speed ($\sim$ 200-250 km s$^{-1}$) ejection of the circumstellar material around V Hya every $\sim$8.5 years which is associated with the periastron passage of a binary companion in an eccentric orbit. 
The first detection of V Hya in the UV was made with IUE \citep{Barnbaum95}. V Hya was subsequently observed and
detected multiple times with GALEX \citep{Sahai08, Montez17}.

%%%%%%%%%%%%%%%%%%%%%%%%%
\section{Observations}\label{Observations}
%%%%%%%%%%%%%%%%%%%%%%%%%

The observations of the ground-state fine structure of the CI($^3P_1$ - $^3P_0$) line at 492.16GHz were performed with the Swedish Heterodyne Facility Instrument \cite[SHeFI;][]{Vassilev08,Belitsky06} on the APEX\footnote{This publication is based on data acquired with the Atacama Pathfinder Experiment (APEX). APEX is a collaboration between the Max-Planck-Institut fur Radioastronomie, the European Southern Observatory, and the Onsala Space Observatory.} 12-meter telescope \citep{Gusten06} located on Llano Chajnantor in northern Chile in Aug 2017 (project ID: O-0100.F-9317A-2017).
Instrument specifics are listed in Table \ref{T1}.

The measured antenna temperatures are converted to the main-beam temperature using $T_{\rm mb}$ = $T^{\ast}_{\rm A}$/$\eta_{\rm mb}$. 
We used the GILDAS/CLASS\footnote{http://www.iram.fr/IRAMFR/GILDAS/} package to reduce the data. A first-order polynomial baseline was subtracted from the averaged spectra. The uncertainty on the absolute intensity scale is estimated to be 20 $\%$.
Total integration times (on+off) are 240 min and 65 min for omi Cet and V Hya, respectively.

We note that frequency shifts of the two isotope $^{13}$CI($^3P_1$-$^3P_0$) hyperfine lines with respect to the frequency of the $^{12}$CI($^3P_1$ -$^3P_0$) line are only 0.5 MHz and 3.6 MHz for the F =1/2$-$1/2 and 3/2$-$1/2 components, respectively, indicating that we have both $^{12}$C and $^{13}$C contributing to the total flux.

\begin{table}[t]
  \centering
  \setlength{\tabcolsep}{5.0pt}
    \caption{\tiny Frequency, main beam efficiency, $\eta_{\rm mb}$, half power beam width, $\theta_{\rm mb}$, and  the excitation energy of the upper transition level, $E_{\rm up}$.}
  \begin{tabular}{@{} ccccccc@{}}
\hline
Trans. & Tel. & Freq. [ GHz]& $\eta_{\rm mb}$ & $\theta_{\rm mb}['']$ & $E_{\rm up}$ [K] \\
\hline
CI(1-0) & APEX & 492.16 & 0.60 & 12.67 & 23.62
 \end{tabular}
%\tablefoot{}
 \label{T1}
\end{table}

%%%%%%%%%%%%%%%%%%%%%%%%%
\section{Results and Discussion}\label{RD}
%%%%%%%%%%%%%%%%%%%%%%%%%

%%%%%%%%%%%%%%%%%%%%%%%%%
\subsection{omi Cet}
%%%%%%%%%%%%%%%%%%%%%%%%%

We present the spectrum of omi Cet in Fig.~\ref{OmiCet} and summarise the results in Table \ref{T2}. In the following subsections we describe the radiative transfer (RT) models that we run to determine the circumstellar CI abundance.

%Atomic carbon has three fine-structure levels in the ground state, $^3P_0$, $^3P_1$, and $^3P_2$, which lie 0, 23, and 62 K, above the ground state, respectively.  The 492 GHz line corresponds to the 3P1-3P0 transition.

\begin{table}[t]
  \centering
  \setlength{\tabcolsep}{4.0pt}
    \caption{The CI Observational results for omi Cet.}
  \begin{tabular}{@{} ccccccc@{}}
%    \toprule
\hline
Star &  $V_{\rm LSR}$ &  $V_{\rm c}$ & $I_{\rm CI}$& $T_{\rm peak (MB)}$ & $\Delta V$  \\
        & [km s$^{-1}$]    & [km s$^{-1}$] &[K km s$^{-1}$] & [K] & [km s$^{-1}$]  \\
\hline
omi Cet &  47.2 &  43.4$\pm$0.6  & 0.55$\pm$0.07 & 0.054 & 9.5$\pm$1.3
\\
   \hline
 \end{tabular}
\tablefoot{\tiny The stellar velocity $V_{\rm LSR}$ was taken from \cite{Khouri16}. The central velocity, $V_{\rm c}$, velocity-integrated line intensity, $I_{\rm CI}$, the peak line temperature, $T_{\rm CI(MB)}$, and the FWHM linewidth, $\Delta$V, were found by fitting a Gaussian line profile to the data.}
 \label{T2}
\end{table}

%%===========================================================================
\begin{figure}[t]
\centering
 \includegraphics[width=.5\textwidth]{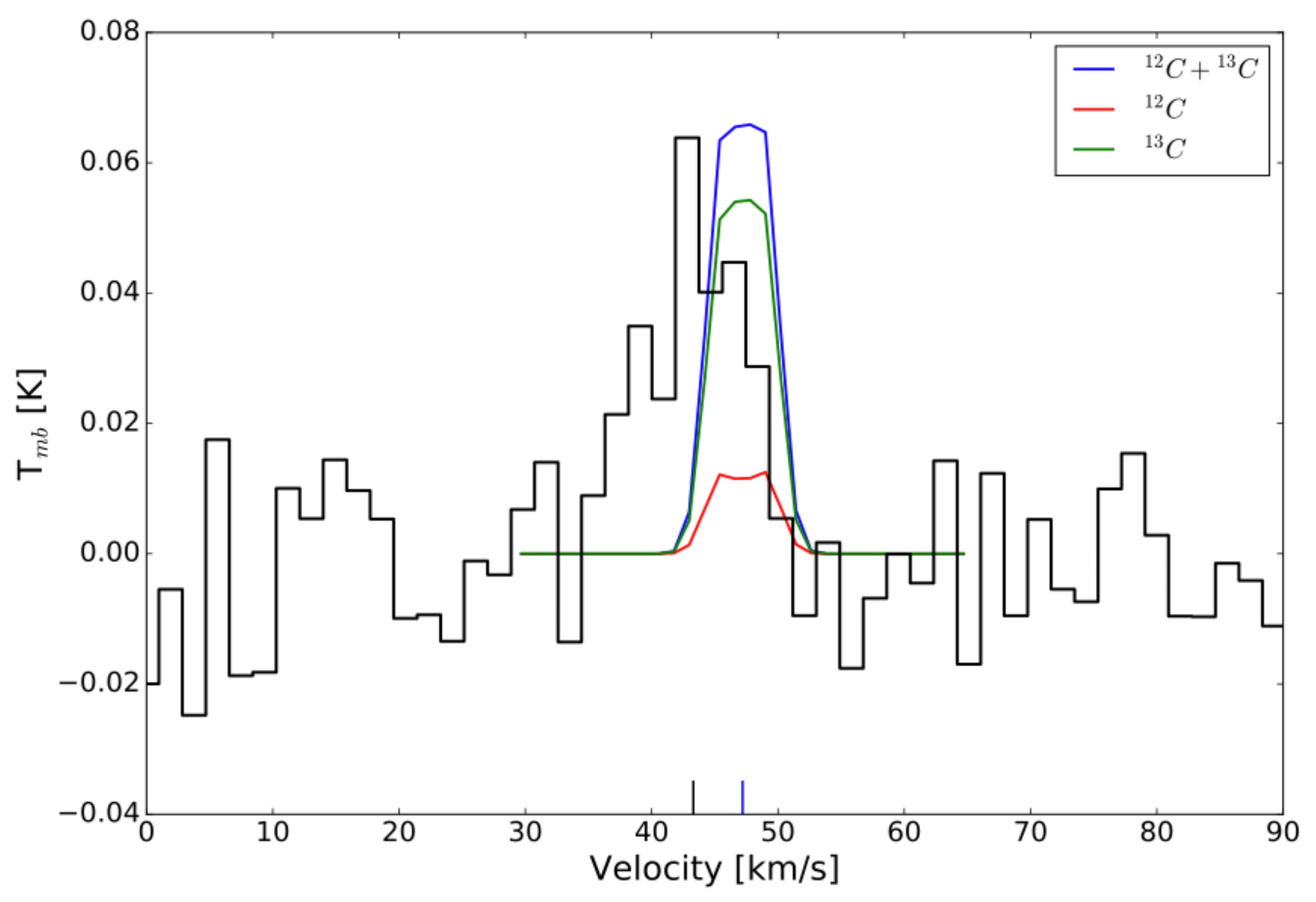}
 \caption[]{\label{OmiCet} 
 \tiny CI emission towards omi Cet at 1.8 km/s velocity resolution \emph{(black)}. The stellar v$_{\rm LSR}$ (47.2km/s) and the spectrum peaks v$_{\rm LSR}$  (43.4km/s) are indicated as vertical markers on the x-axis in blue and black, respectively. The results of RT modelling of both $^{12}$C and $^{13}$C isotopes are also shown in \emph{red} and \emph{green}, respectively. The \emph{blue} profile indicates the total amount of C.}
  \label{OmiCet}  
\end{figure}
%%===========================================================================

%%%%%%%%%%%%%%%%%%%%%%%%%%%%%%%%%%%%%%%%%%
\subsubsection{The circumstellar model}\label{CSE-model}
%%%%%%%%%%%%%%%%%%%%%%%%%%%%%%%%%%%%%%%%%%

To model the envelope around omi Cet, we assume a uniformly expanding spherical envelope which is formed due to a constant mass-loss. 
We adopt the physical properties of the CSE from CO RT modelling results obtained by RS01; see Table 3 in that paper. 
Using the given CSE properties, the updated CO molecular data, and the updated distance of 92 pc for omi Cet,
we could reasonably model the CO($J$=2-1,3-2,4-3) emission lines observed by JCMT.
Although the model reasonably reproduces the CO observations, it is not possible to constrain the complex outflow around omi Cet based on the single-dish observations; it is therefore a crude approximation of the CSE properties. 

A non-local thermodynamic equilibrium (non-LTE) RT code based on the Monte Carlo program (mcp) \citep[see e.g.][]{Bernes79,Schier01} was used to analyse the circumstellar CI emission. 
The CI is assumed to be excited by collision with H$_2$ molecules and through radiation from the central star, the dust and the cosmic microwave background. The collisional data are taken from \cite{Schroder91}. They cover temperatures from 10 to 500 K.

To derive the CI abundance distribution through the envelope, we used an extended version of the publicly available circumstellar envelope chemical model code\footnote{http://udfa.ajmarkwick.net/index.php?mode=downloads} \citep{McElroy13}. 
The extended version includes $^{13}$C and $^{18}$O isotopes and the molecules containing these isotopes in addition to a more extended chemical network (Saberi et al. in prep.).
The result of the chemical modelling is presented in the Appendix. We used the $^{12}$C and $^{13}$C abundance distributions that are shown in Fig.~\ref{Chemi} as input files in the mcp.
% \LEt{Please spell out all acronyms the first time they appear in the paper, followed by the abbreviation in parentheses, both in the abstract and again in the main text. After that, please only use the abbreviation. See A and A language guide Section 5.2.4 www.aanda.org/language-editing}

Results of the RT modelling for both C isotopes and the integrated flux of both isotopes are presented in Fig.~\ref{OmiCet}. The flux is dominated by $^{13}\rm CI$ line emission since it originates closer to the star where $^{12}$CO is more efficiently self-shielded while the less abundant $^{13}$CO would be dissociated in the more inner region \citep{Lee84}.

The model produces a narrower line profile than observed in the spectrum. The narrow width of the profile is due to the low expansion velocity 2.5 km s$^{-1}$ of the CSE that was assumed in the RS01 model. There is also a shift ($\sim 4$ km s$^{-1}$) between the central peak of the spectrum and the model. 
We cannot explain the velocity shift under the assumption of an external UV field impacting on a regular circumstellar envelope which leads a spherically symmetric CI distribution, centred around the star, and hence around the stellar v$_{\rm LSR}$.

%%%%%%%%%%%%%%%%%%%%%%%%%%%%%%%%%%%%%%%%%%
\subsubsection{Constraining the CI emitting region}
%%%%%%%%%%%%%%%%%%%%%%%%%%%%%%%%%%%%%%%%%%

\begin{figure}[t]
% \centering
 \includegraphics[width=90mm]{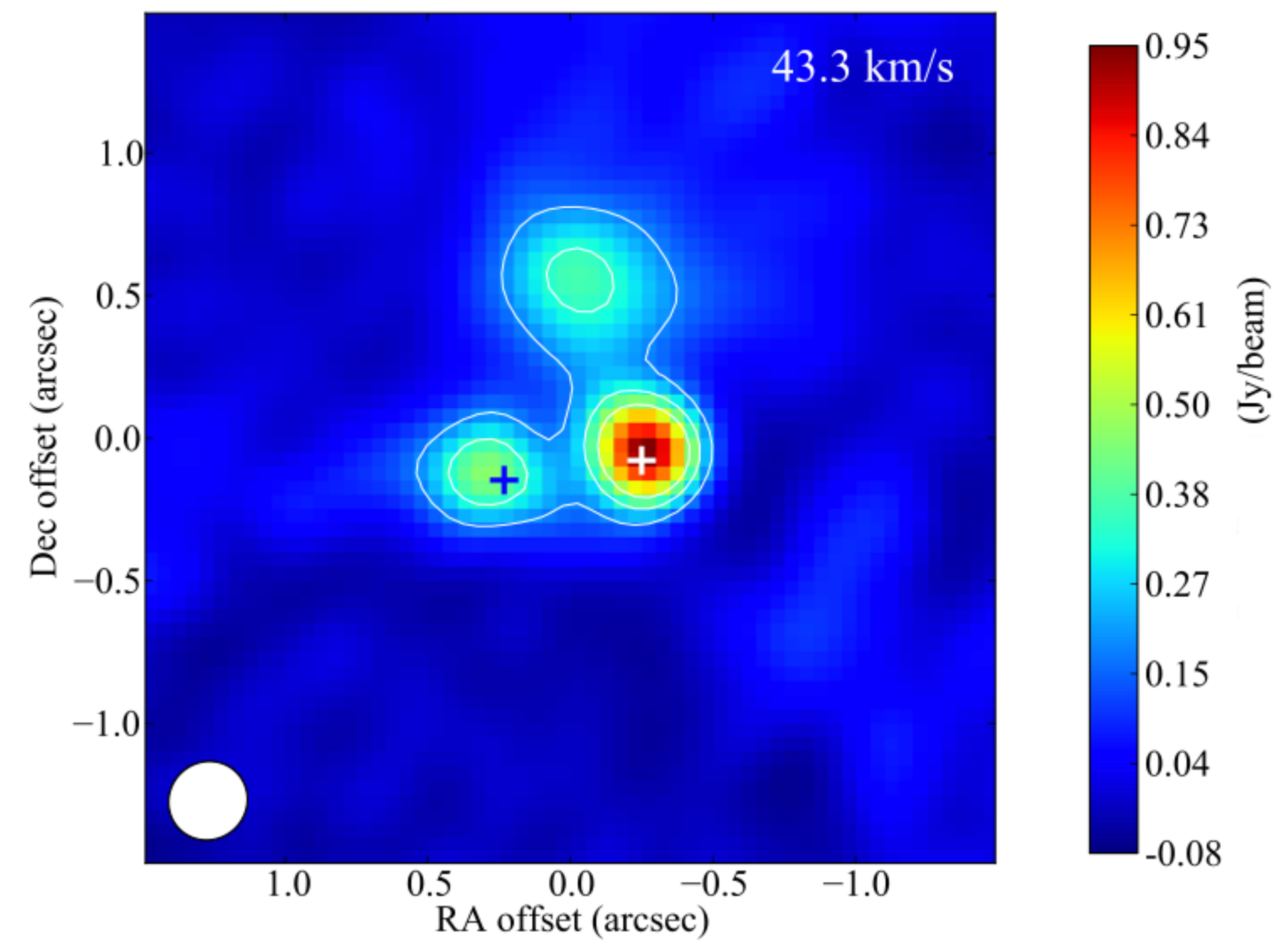}
 \caption[]{\tiny ALMA observations of CO(3-2) towards the Mira AB system at v$_{\rm lsr}$=43.3 km s$^{-1}$. The white and blue plus symbols show the positions of omi Cet (Mira A) and VZ Cet (Mira B), respectively. The contours are overlaid on the image with 20, 35, and 50$\%$ of the peak value. The observed CI emission likely arises from the region near Mira B. The beam is shown in the lower left corner.}
\label{OmiCet-ALMA}  
\end{figure}

The observed UV fluxes towards the symbiotic binary suggest that the measured CI spectrum could arise from a more compact region near the binary companion.
We combined the CO(3-2) ALMA observations from projects 2012.1.00524.S (PI: Ramstedt) and 2013.1.00047.S (PI: Planesas) to reach an angular resolution of 0.28$''$. 
We show the CO(3-2) line emission map at v$_{\rm LSR}$=43.3km/s, the closest to the CI peak velocity, in Fig. \ref{OmiCet-ALMA}. 
 As shown, there are three components that peak around Mira A, Mira B, and the north part of the Mira AB system in the CO(3-2) emission.
If the total CI flux was coming from Mira A, we would not expect to observe such a velocity shift between the CI peak emission and the stellar velocity as we discussed in the Sect. \ref{CSE-model}. 
There are no known sources of UV- or shock-dissociation near the clump in the north part that cause the enhancement of the CI emission. Therefore, the most likely hypothesis is that CI arises from the component near Mira B due to the strong UV emission.
The velocity shift could then be the result of the orbital velocity of Mira B around Mira A. Although the orbital velocity of Mira B is not yet accurately known, based on the orbital parameters \citep{Prieur02} it is of the order of 6 km/s.

%%===========================================================================
%%===========================================================================

%%%%%%%%%%%%%%%%%%%%%%%%%
\subsubsection{The CI abundance in the postulated emitting region}\label{emitting region}
%%%%%%%%%%%%%%%%%%%%%%%%%

We derive the CI abundance assuming it comes from the component around Mira B seen in Fig. \ref{OmiCet-ALMA}. This component has a radius $\sim$ 0.2$''$ = 2.67 $\times 10^{14}$ cm.
The publicly available non-LTE RT code RADEX\footnote{http://var.sron.nl/radex/radex.php} is used to model the CI emission. 

The input parameters used in RADEX are listed in Table \ref{T3}. 
The H$_2$ number density is calculated assuming a constant mass-loss rate and constant expansion velocity as in \cite{Schier00}. We assumed the temperature profile T(r) = 150 (10$^{15}$/r) as given by RS01.
Here we assumed the updated distance 92 pc reported by \cite{vanLeeuwen07} since using the correct distance is crucial in calculating the column density.
We consider constant H$_2$ number density and kinetic temperature profiles in the constrained region. These are calculated at the Mira B position r = 60 AU = 8.97 $\times10^{14}$ cm from omi Cet.
The line width is estimated by fitting a Gaussian profile to the observed spectrum.
The only free parameter in this modelling is the CI column density. 

In RADEX, the calculated radiation temperature of the spectral line T$_{\rm R}$ is equivalent to the main beam antenna temperature divided by the dilution factor for an unresolved source. With a telescope beam size of $\sim$13$''$  and assuming an emission region of 0.4$''$, we get a dilution factor of $(0.4/13)^2\sim 0.001$. This leads to a corrected main beam temperature of 0.054 / 0.001 = 54 K.
To reproduce this temperature, a CI column density of $1.1\times10^{19}$ cm$^{-2}$ is needed for the component near to Mira B. 
We note that the radiation temperature T$_{\rm R}$ is not very sensitive to the H$_2$ density and the kinetic temperature. 
T$_{\rm R}$ does not change more than 10 $\%$ over ranges of $1\times10^6<n_{\rm H_2}<1\times10^{12}$ cm$^{-3}$ and $100 <T_{\rm kin}<500$ K.

The CI/H$_2$ ratio strongly depends on the true size and the H$_2$ number density of the emitting region. Since the circumstellar material of omi Cet is accreted onto and heated by the binary companion, we would expect a higher H$_2$ density and temperature in the clump compared to the values that were derived from the smooth wind model.
 \cite{Ireland07} infer an overdensity around Mira B of a factor of 25-100 compared to the smooth wind model.
If we assume the H$_2$ density is $1.9\times10^8$ cm$^{-3}$ in the clump (higher by a factor of 100 compared to the smooth wind model), then our measurement implies a fractional abundance CI/H$_2$ $\sim1.1\times 10^{-4}$ in this location.
However, to accurately determine the CI/H$_2$ ratio we need to constrain the true size of the emitting region, requiring high-angular-resolution observations of the CI line emission. Additionally, high-angular-resolution observations of multiple transitions of a collisionally excited molecule like CO would allow a better H$_2$ density estimate in the CI-emitting region.

\begin{table}[t]
  \centering
  \setlength{\tabcolsep}{3.5pt}
    \caption{ The input parameters used in RADEX to determine the CI abundance towards omi Cet.}
  \begin{tabular}{@{}ccccc@{}}
\hline
T$_{\rm back.}$. &  T$_{\rm Kin}$  &  H$_2$ density & Line width & CI column density \\
\hline
[K]& [K] & [ cm$^{-3}$]  & [km s$^{-1}$] & [ cm$^{-2}$]\\
\hline
2.73 &  170 &  1.9E6  & 9.5 & 1.1E19
\\
 \hline
 \end{tabular}
%\tablefoot{}
 \label{T3}
\end{table}

%%%%%%%%%%%%%%%%%%%%%%%%%
\subsubsection{The CI/CO ratio}
%%%%%%%%%%%%%%%%%%%%%%%%%

To derive the CI/CO ratio, we need to know the size and the H$_2$ density of the emitting region. Assuming the smooth wind model without enhancement of the H$_2$ density in the clump and the abundance ratio CO/H$_2$ $\sim$ 5 $\times 10^{-4}$ (RS01), we derive a CI/CO $\sim$ 20 which is unrealistically high. If we assume the enhancement of H$_2$ density by a factor of 100 as was explained in Sect. \ref{emitting region} ,we find a ratio CI/CO $\sim$ 0.2.
This ratio is similar to the CI/CO $\sim$ 0.2-0.5 reported in the detached-shell of R Scl \citep{Olofsson15} and the ratio of $\sim$ 0.3 reported for V Hya \citep{Knapp00}, and higher than the ratio $\sim$ 0.02 reported for IRC+10216 \citep{Young97}.

It is worth mentioning that a strong detection of CI was reported in the inner part of the O-rich supergiant star $\alpha$ Ori (Betelgeuse) \citep{Huggins94, vanderveen98}. These authors interpret the high observed ratio of CI/CO $\sim$ 5 as being due to the presence of a chromosphere and therefore extra UV radiation.
Similarly, the CI emission towards omi Cet likely arises from the inner region near its binary companion, showing the extra UV-dissociation in the inner CSE around the companion of an AGB star.

%%%%%%%%%%%%%%%%%%%%%%%%%
\subsection{V Hya}\label{VHyasect}
%%%%%%%%%%%%%%%%%%%%%%%%%

The observed spectrum towards V Hya is presented and discussed in the Appendix Fig.~\ref{VHya}.
Our detection of CI for V Hya is consistent with the previous detection reported by \cite{Knapp00}. 
%We discussed the observational results in more details in the Appendix \ref{VHyaS}.

%%%%%%%%%%%%%%%%%%%%%%%%%%
\section{Conclusion}\label{Conclusion}
%%%%%%%%%%%%%%%%%%%%%%%%%%

In this letter, we report the CI line emissions from omi Cet (M-type) and V Hya (C-type) AGB stars. 
omi Cet is the first O-rich AGB star with a CI detection. 
The CI column density of omi Cet is estimated to be $\sim$ 1.1 $\times$ 10$^{19}$ cm$^{-2}$ if the emission arises from a compact region near the hot secondary star, Mira B. In that case, the UV emission from Mira B and/or from the accretion of matter from the wind of Mira A onto Mira B is the likely cause for the observed CI enhancement. 
On the other hand, the observed flux is consistent with CI in a shell produced by CO dissociation from the ISRF. However this model does not correctly explain the observed line width and the velocity shift between the peak of CI emission and the stellar velocity. 
Since there could be other kinematic effects in such a complex envelope however, we cannot confidently rule out the external UV field.
Only higher-angular-resolution maps will be able to determine the real CI distribution and thus the origin of the CI enhancement.

Definite detections of CI were previously reported in IRC+10216 and R Scl, two carbon-rich AGB stars.
Our observations bring the number of CI detection of AGB stars to four in total and the number of evolved stars (including transition objects) to nine \cite[][and references therein]{Knapp00}. 
 
The spatial abundance distribution of circumstellar products of photodissociation and ionisation can provide insight into the relevant sources of UV radiation, by probing the effects of (unidentified) hot binary companions and stellar chromospheric activity of AGB stars.

%%%%%%%%%%%%%%%%%%%%%%%%%%
\begin{acknowledgements}
%%%%%%%%%%%%%%%%%%%%%%%%%%

This work was supported by ERC consolidator grant 614264. EDB acknowledges financial support from the Swedish National Space Board. This paper makes use of the following ALMA data: ADS/JAO.ALMA\#2012.1.00524.S and ADS/JAO.ALMA\#2013.1.00047.S. ALMA is a partnership of ESO (representing its member states), NSF (USA) and NINS (Japan), together with NRC (Canada), NSC and ASIAA (Taiwan), and KASI (Republic of Korea), in cooperation with the Republic of Chile. The Joint ALMA Observatory is operated by ESO, AUI/NRAO and NAOJ. We are grateful to the anonymous referee for insightful comments and suggestions that improved the manuscript.

\end{acknowledgements}

% for the bibliography, at the end
\bibliographystyle{aa} % style aa.bst
\bibliography{refrences} % your references Yourfile.bib
%\documentclass[bibyear]{aa}

%%%%%%%%%%%%%%%%%%%%%%%%%%%%%%%%%
\begin{appendix} 
%%%%%%%%%%%%%%%%%%%%%%%%%%%%%%%%%
\section{Chemical modelling results}

Here we present the results of the chemical modelling of omi Cet. 
The code assumes a spherically symmetric envelope which is formed due to a constant mass-loss rate 2.5 $\times$ 10$^{-7}$ M$_\odot$ yr$^{-1}$. The envelop expands with a constant expansion velocity 2.5 km s$^{-1}$. The ISRF is the only UV radiation field which penetrates through the envelope from the outside. We adopt the stellar luminosity and the temperature profile of the CSE from RS01 model.
We assumed the initial fractional abundances of $^{12}$CO/H$_2$=5$\times$10$^{-4}$ reported for omi Cet (RS01), the $^{13}$CO/H$_2$=5$\times$10$^{-5}$ based on the isotopic ratio reported by \cite{Hinkle16}, and the H$^{12}$CN/H$_2$=1$\times$10$^{-7}$ which is the average ratio reported for M-type AGB stars \citep{Schier13}. We do not expect a significant contribution of other C-bearing molecules for an M-type AGB star in the model. 
Figure \ref{Chemi} shows the fractional abundance distribution of some circumstellar species through the envelope. We use the $^{12}$C and $^{13}$C distribution profiles as the input in our RT model.

%%===========================================================================
\begin{figure}[]
\centering
 \includegraphics[width=.46\textwidth]{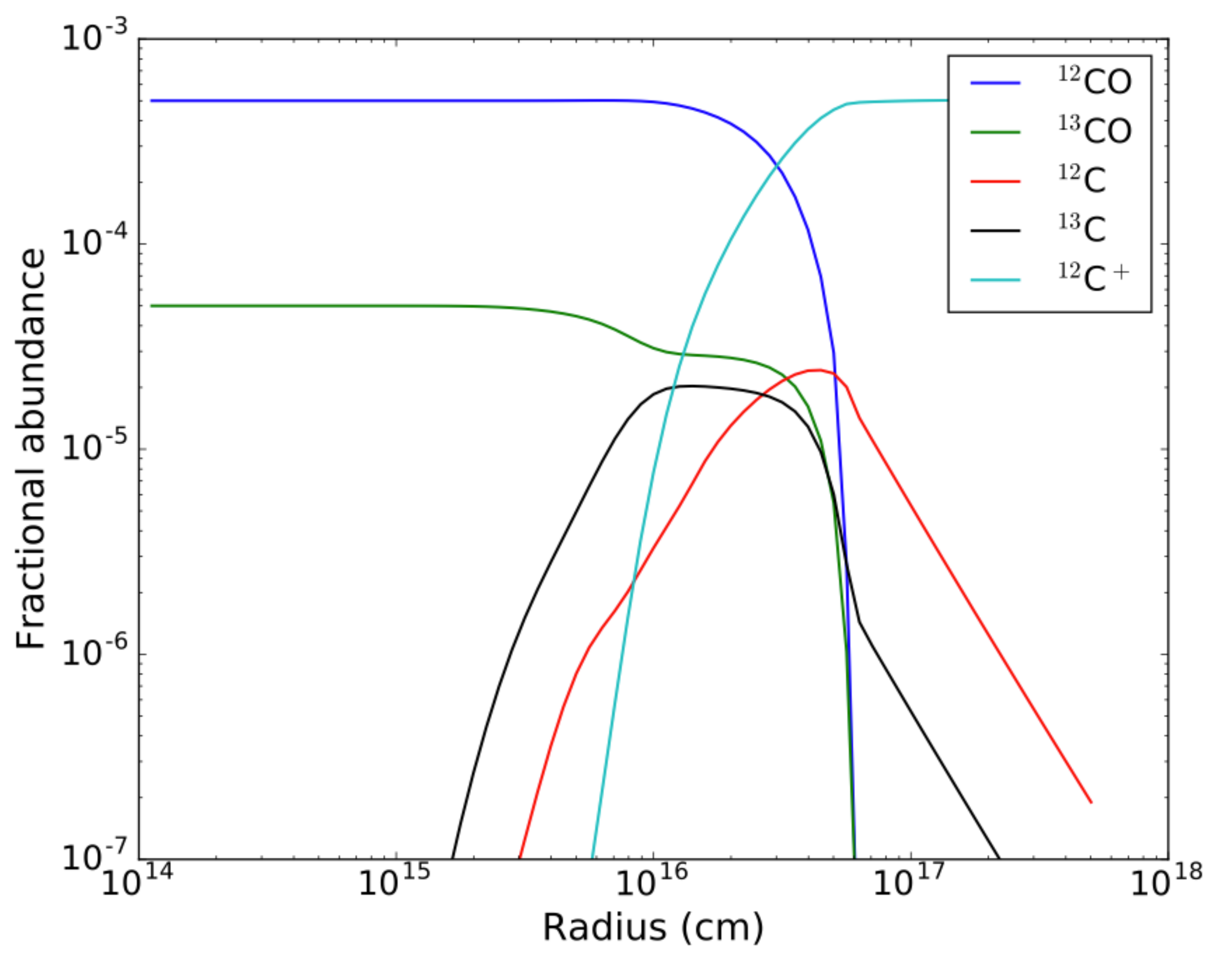}
 \caption[]{\label{Chemi} 
  The fractional abundance distribution of the circumstellar species of the Omi Cet.}
  \label{Chemi}  
\end{figure}
%%===========================================================================

\section{V Hya spectrum}\label{VHyaS}

\begin{figure*}[t]
  \includegraphics[width=\textwidth,height=6.2cm]{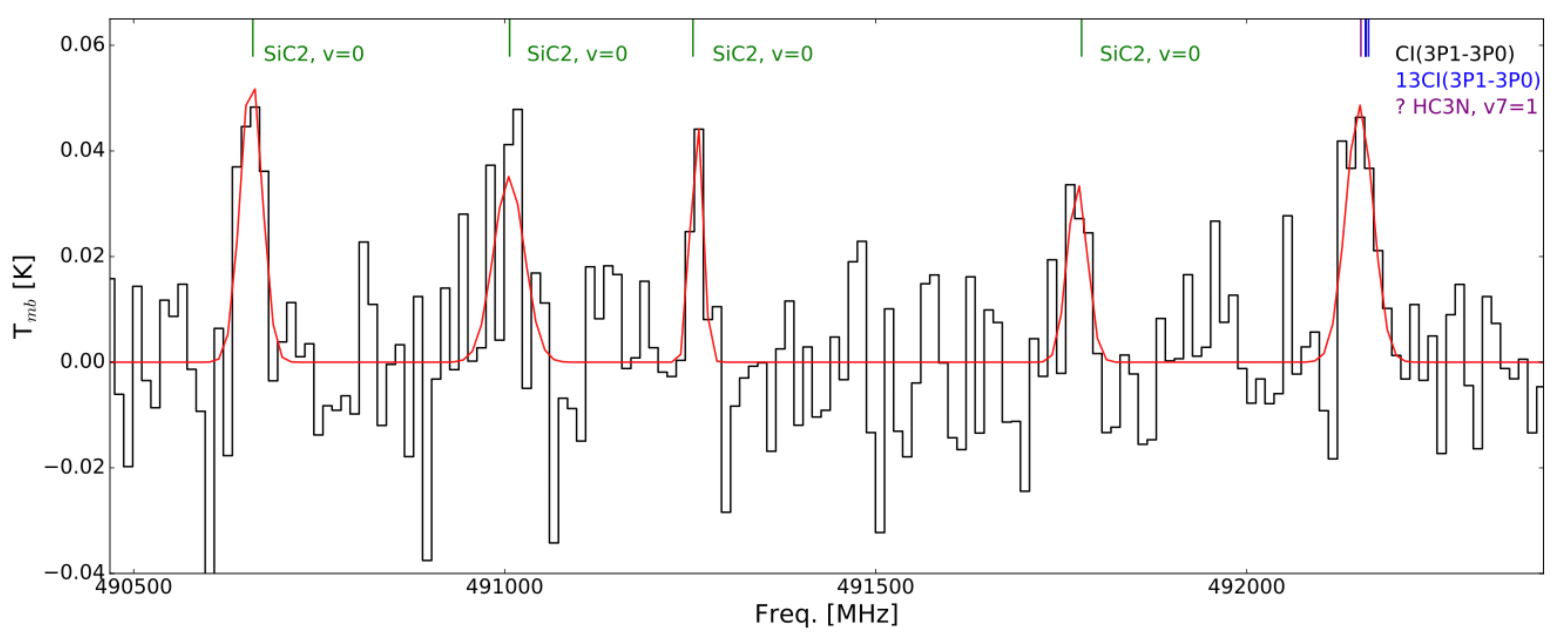}
  \caption{ APEX observations towards V Hya at v$_{\rm lsr}$= -17 km s$^{-1}$ with the spectral resolution 12.21 MHz (7.4 km s$^{-1}$) \emph{(black)}. 
  The red line indicates a Gaussian fit to potential line detections.}
  \label{VHya}  
\end{figure*}

Figure \ref{VHya} shows the observed spectrum of V Hya.
Although our detection of CI for V Hya is consistent with the previous detection reported by \cite{Knapp00}, the CI line emission could potentially be contaminated by HC$_3$N ($\nu_7$=1) vibrationally excited emission. The vibrationally excited HC$_3$N has been detected for the proto-planetary nebula CRL618 by \cite{Wyrowski03}.
A detection of circumstellar HC$_3$N ground state emissions around V Hya was previously reported by \cite{Knapp97}.
In addition, preliminary chemical modelling results show a large enhancement of HC$_3$N in the inner CSE by the internal UV radiation (Saberi et al. in prep).
  
As shown in Fig.~\ref{VHya}, we potentially identify detection of several SiC$_2(\nu=0)$ emission lines in the V Hya spectrum.
\cite{Sarre00} have presented the SiC$_2$ absorption bands in the upper atmosphere of V Hya.
We cannot differentiate the possible contributions of different lines because of the low resolution of the spectrum. Therefore, we did not analyse the V Hya spectrum any further.

%%===========================================================================

%%===========================================================================

 \end{appendix}

\end{document}